\documentclass[iop]{emulateapj}
\usepackage{comment,threeparttable}

\newcommand{\msun}{M$_\sun$}
\newcommand{\msunyear}{M$_\sun$yr$^{-1}$}
\newcommand{\igyr}{Gyr$^{-1}$}

\newcommand{\bdrop}{$B_{435}-$dropout}
\newcommand{\bbdrop}{$B-$dropout}

\newcommand{\fourstar}{{\sc FourStar}}
\newcommand{\jo}{$J_1$}
\newcommand{\jt}{$J_2$}
\newcommand{\jh}{$J_3$}
\newcommand{\hs}{$H_s$}
\newcommand{\hl}{$H_l$}
\newcommand{\ks}{$Ks$}

\newcommand{\ms}{$M$}
\newcommand{\vj}{$V$-$J$}
\newcommand{\uvj}{$UVJ$}
\newcommand{\uv}{$U$-$V$}

\newcommand{\numgals}{57}
\newcommand{\numqu}{26}
\newcommand{\numsf}{8}
\newcommand{\numdu}{23}
\newcommand{\fracqu}{$0.46^{+0.06+0.10}_{-0.06-0.17}$}
\newcommand{\fracsa}{$0.54^{+0.08+0.17}_{-0.08-0.10}$}
\newcommand{\fracsf}{$0.14^{+0.03+0.10}_{-0.03-0.04}$}
\newcommand{\fracdu}{$0.40^{+0.06+0.07}_{-0.06-0.05}$}
\newcommand{\percqu}{$46^{+6+10}_{-6-17}$\%}
\newcommand{\percsa}{$54^{+8+17}_{-8-10}$\%}
\newcommand{\percsf}{$14^{+3+10}_{-3-4}$\%}
\newcommand{\percdu}{$40^{+6+7}_{-6-5}$\%}
\newcommand{\muv}{$M_{1700}=-18.05\pm0.37$}
\newcommand{\eazy}{{\tt EAZY}}

\shorttitle{ZFOURGE: The massive galaxy population at $z=3-4$}
\shortauthors{L.~Spitler et al.}

\begin{document}

\title{Exploring the $z=3-4$ massive galaxy population with ZFOURGE:\\
the prevalence of dusty and quiescent galaxies\altaffilmark{1}}


\author{Lee~R.~Spitler\altaffilmark{2,3,4},
Caroline~M.~S.~Straatman\altaffilmark{5},
Ivo~Labb\'e\altaffilmark{5},
Karl~Glazebrook\altaffilmark{4},
Kim-Vy~H.~Tran\altaffilmark{6},
Glenn~G.~Kacprzak\altaffilmark{4,7},
Ryan~F.~Quadri\altaffilmark{8,9},
Casey~Papovich\altaffilmark{6},
S.~Eric~Persson\altaffilmark{8},
Pieter~van~Dokkum\altaffilmark{10},
Rebecca~Allen\altaffilmark{3,4},
Lalitwadee~Kawinwanichakij\altaffilmark{6},
Daniel~D.~Kelson\altaffilmark{8},
Patrick~J.~McCarthy\altaffilmark{8},
Nicola~Mehrtens\altaffilmark{6}
Andrew~J.~Monson\altaffilmark{8},
Themiya~Nanayakkara\altaffilmark{4},
Glen~Rees\altaffilmark{2,11},
Vithal~Tilvi\altaffilmark{6},
Adam~R.~Tomczak\altaffilmark{6}
}

\email{lee.spitler--at--mq.edu.au}
\altaffiltext{1}{Based on data gathered with the 6.5 meter Magellan Telescopes located at Las Campanas 
Observatory, Chile.}
\altaffiltext{2}{Department of Physics \& Astronomy, Macquarie University, Sydney, NSW 2109, Australia}
\altaffiltext{3}{Australian Astronomical Observatory, PO Box 296 Epping, NSW 1710, Australia}
\altaffiltext{4}{Centre for Astrophysics \& Supercomputing, Swinburne University, Hawthorn, VIC 3122, Australia}
\altaffiltext{5}{Leiden Observatory, Leiden University, P.O. Box 9513, 2300 RA Leiden, The Netherlands}
\altaffiltext{6}{George P. and Cynthia W. Mitchell Institute for Fundamental Physics and Astronomy, Department of Physics and Astronomy, Texas A\&M University, College Station, TX 77843, USA}
\altaffiltext{7}{Australian Research Council Super Science Fellow}
\altaffiltext{8}{Carnegie Observatories, Pasadena, CA 91101, USA}
\altaffiltext{9}{Hubble Fellow}
\altaffiltext{10}{Department of Astronomy, Yale University, New Haven, CT 06520, USA}
\altaffiltext{11}{Australia Telescope National Facility, CSIRO Astronomy \& Space Science, PO Box 76, Epping, NSW 1710, Australia}
 
\begin{abstract}
Our understanding of the redshift $z>3$ galaxy population relies largely on samples selected using the popular ``dropout'' technique, typically consisting of UV-bright galaxies with blue colors and prominent Lyman breaks.  As it is currently unknown if these galaxies are representative of the massive galaxy population, we here use the FourStar Galaxy Evolution (ZFOURGE) Survey to create a stellar mass-limited sample at $z=3-4$. Uniquely, ZFOURGE uses deep near-infrared medium-bandwidth filters to derive accurate photometric redshifts and stellar population properties. The mass-complete sample consists of 57 galaxies with log~\ms~$>10.6$, reaching below $M^{\star}$ at $z=3-4$.

On average, the massive $z=3-4$ galaxies are extremely faint in the observed optical with median $R_{tot}^{AB}=27.48\pm0.41$ (restframe \muv{}). They lie far below the UV luminosity-stellar mass relation for Lyman break galaxies and are about $\sim100\times$ fainter at the same mass.  The massive galaxies are red ($R-Ks_{AB}=3.9\pm0.2$; restframe UV-slope $\beta=-0.2\pm0.3$) likely from dust or old stellar ages. We classify the galaxy SEDs by their restframe \uv{} and \vj{} colors and find a diverse population: \percqu{} of the massive galaxies are quiescent, \percdu{} are dusty star-forming galaxies, and only \percsf{} resemble luminous blue star forming Lyman break galaxies. This study clearly demonstrates an inherent diversity among massive galaxies at higher redshift than previously known.  Furthermore, we uncover a reservoir of dusty star-forming galaxies with $4\times$ lower specific star-formation rates compared to submillimeter-selected starbursts at $z>3$. With $5\times$ higher numbers, the dusty galaxies may represent a more typical mode of star formation compared to submillimeter-bright starbursts.
\end{abstract}


\section{Introduction}\label{intro}

\begin{figure*}
\includegraphics[width=\textwidth]{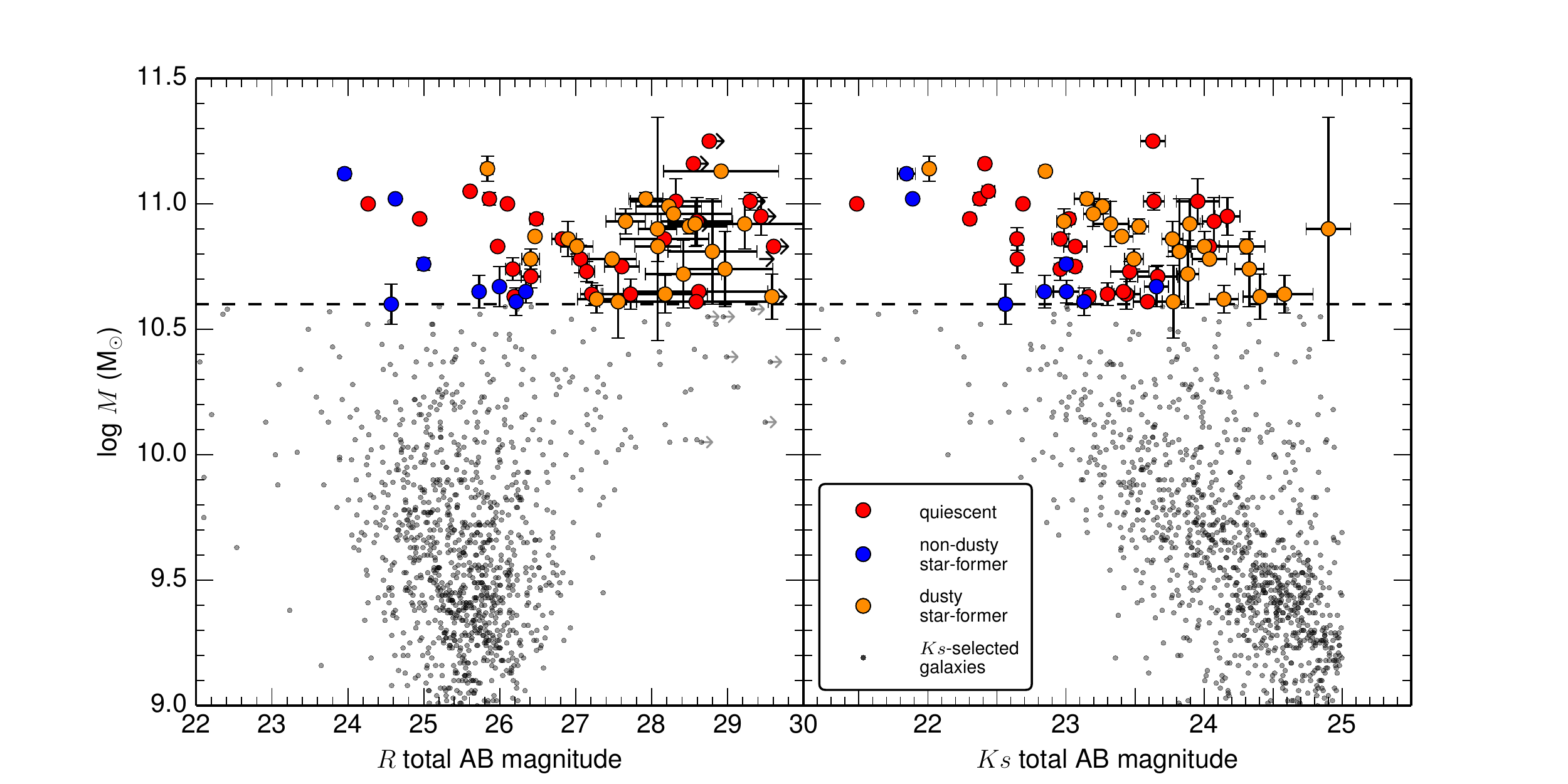}
\caption{
Observed magnitude versus stellar mass of our ZFOURGE $z=3-4$ galaxy sample. Colored points are in our mass-limited sample, while the black points are all \ks{}-selected $z=3-4$ galaxies. Colors reflect subgroups within the restframe \uvj{} color-color parameter space, as shown in Fig.~\ref{fig_uvj}. At observed $R$-band ($3\sigma$ limits shown), the massive galaxies in our sample are significantly fainter (median $R_{tot}=27.48\pm0.41$) than the lower-mass ZFOURGE galaxies ($R_{tot}=25.63\pm0.02$), while at $Ks$-band the mass-limited sample is more luminous.  Error bars along the top reflect the median uncertainties on the measured quantities for our mass-limited sample.}
\label{fig_summary}
\end{figure*}

To date, typical galaxy population studies at $z>3$ utilized the ``dropout'' technique to efficiently identify high-redshift galaxies with steep Lyman breaks \citep[e.g. ][]{steidel_ly_2000}. While this technique has been used successfully to push our understanding of galaxies out to $z\sim10-12$ \citep{bouwens_candidate_2011,ellis_abundance_2013,coe_clash:_2013}, the results that can be drawn from these studies apply only to the subset of restframe UV-bright star-forming galaxies with prominent Lyman breaks and are not necessarily applicable to the general galaxy population at $z>3$ \citep[e.g. ][]{cooke_lyman_2014,wyithe_predicted_2013}. It is therefore unclear how dropout galaxies relate to other galaxies at $z>3$, including submillimeter \citep{smail_nature_2002,chapman_redshift_2005} and quiescent galaxies \citep[e.g. ][]{straatman_substantial_2014}.

To advance our understanding of galaxies at redshifts $z>3$, a complete census of massive galaxies must obtained. A broad range of spectral features must therefore be available to determine photometric redshifts (e.g. the Lyman, Balmer, $4000$~\AA{} breaks, 1.6~$\mu$m ``bump''). This ensures the identification is not limited to any particular spectral feature.

Key to conducting a full galaxy census is to observe redward of the Balmer/$4000$~\AA{} breaks, which is necessary for identifying galaxies with low levels of star formation and/or significant dust content. Unfortunately, these spectral breaks fall into the observed near-infrared, which is a very challenging observational regime for ground-based, small area near-infrared cameras.

Broadband infrared imaging has enabled some progress \citep[e.g. ][]{chen_discovery_2004,wiklind_population_2008,fontana_fraction_2009,mancini_searching_2009}. More recently, wide area but relatively shallow near-IR imaging allowed the first characterization of the massive galaxy population at $z=3-4$ \citep{marchesini_most_2010,caputi_nature_2012,ilbert_mass_2013,muzzin_evolution_2013,stefanon_what_2013}. These studies have shown that this population is not just made up of Lyman break galaxies, but includes many dusty and quiescent galaxies as well.

In this Letter, we put this result on solid footing, using much deeper near-IR data that allows us to extend to lower masses (below $M^{\star}$ at $z=3$) and access ``normal'' galaxies. We utilize new deep observations from the FourStar Galaxy Evolution (ZFOURGE) survey\footnote{{\url http://zfourge.tamu.edu}} (Labb{\'e} et al., in prep.), which include near-infrared medium-bands that allows us to precisely sample the SEDs and obtain improved photometric redshifts and stellar population estimates.

In the following we adopt the AB magnitude system and a cosmology with $H_0 = 70$ km s$^{-1}$ Mpc$^{-1}$, $\Omega_m=0.3$, and $\Omega_{\Lambda}=0.7$. Scatter on quantities are from bootstrap analysis.

\section{Observations and data}\label{observations}

\begin{table*}
\begin{center}
\begin{threeparttable}
\caption{Median properties of mass-selected galaxies at $\mathrm{3\leq\mathit{z}<4}$}
\begin{tabular}{l l r r r r r r r r r r r }
\hline
\hline
 &  & subgroup\tnote{a} &          & log \ms{}         &  $Ks_{tot}$ & $R_{tot}$          & $U$-$V$     & $V$-$J$     & $\beta$\tnote{b,c,e} & $M_{1700}$\tnote{b,e} \\
 &  N  & fraction &     $z_{phot}$ & (\msun{}) & (AB)       & (AB)               &  restframe     & restframe     & UV slope         & (AB)       \\
\hline

all &  57 & \nodata & $3.34$       & $10.83\pm 0.04$     & $23.32\pm 0.14$      & $27.48\pm 0.41$     & $ 1.54 \pm  0.04$ & $ 1.20 \pm  0.11$ & $  -0.2 \pm    0.3$  & $-18.05 \pm  0.37$ \\
    & & &           &            &           &     & $\sigma= 0.25$ & $\sigma= 0.59$ & $\sigma=   1.4$  & $\sigma= 1.92$ \\
    quiescent &  26 & \fracqu & $3.52$       & $10.86\pm 0.06$     & $23.12\pm 0.16$      & $27.25\pm 0.52$     & $ 1.61 \pm  0.09$ & $ 0.85 \pm  0.07$ & $  -0.3 \pm    0.5$  & $-19.43 \pm  0.64$ \\
        & & &           &            &           &     & $\sigma= 0.29$ & $\sigma= 0.23$ & $\sigma=   1.5$  & $\sigma= 1.59$ \\
	star-forming  &  31 & \fracsa & $3.22$       & $10.83\pm 0.04$     & $23.49\pm 0.21$      & $27.65\pm 0.65$     & $ 1.45 \pm  0.08$ & $ 1.54 \pm  0.08$ & $  -0.2 \pm    0.4$  & $-18.01 \pm  0.49$ \\
	   (SF) & & &           &            &           &     & $\sigma= 0.33$ & $\sigma= 0.35$ & $\sigma=   1.4$  & $\sigma= 1.25$ \\
	    non-dusty SF &   8 & \fracsf & $3.21$       & $10.66\pm 0.08$     & $22.92\pm 0.41$      & $25.36\pm 0.57$     & $ 1.07 \pm  0.05$ & $ 0.80 \pm  0.16$ & $  -0.6 \pm    0.3$  & $-20.53 \pm  0.28$ \\
	        & & &           &            &           &     & $\sigma= 0.11$ & $\sigma= 0.37$ & $\sigma=   0.6$  & $\sigma= 0.40$ \\
		dusty SF &  23 & \fracdu & $3.23$       & $10.86\pm 0.04$     & $23.78\pm 0.20$      & $28.17\pm 0.29$     & $ 1.56 \pm  0.06$ & $ 1.65 \pm  0.06$ & $   0.6 \pm    0.6$  & $-17.83 \pm  0.17$ \\
		    & & &           &            &           &     & $\sigma= 0.18$ & $\sigma= 0.19$ & $\sigma=   1.0$  & $\sigma= 0.68$ \\

\end{tabular}
\quad
\begin{tabular}{l r r c r r r c }
\hline
\hline
 & $n$          & $n$\tnote{d}    & SFR\tnote{b}         & sSFR\tnote{b} & log $\rho_{SFR}$\tnote{b,e} & log $\rho_{\ast}$ &  \\
 & (arcmin$^{-2}$) & ($10^{-5}$ Mpc$^{-3}$) & (\msunyear{}) & (\igyr{}) & (\msunyear{} Mpc$^{-3}$) & (\msun{} Mpc$^{-3}$) & AGN \%\tnote{d}\\
\hline

all & 0.144 & $5.1\pm0.7$ & $ 51\pm 51$ & $ 0.78\pm0.62$ & -2.32 & 6.59 & $21\pm7$ \\
quiescent & 0.066 & $2.3\pm0.6$ & $ 10\pm  3$ & $ 0.16\pm0.03$ & -3.64 & 6.27 & $23\pm10$ \\
all SF & 0.078 & $2.8\pm0.5$ & $181\pm 59$ & $ 2.61\pm0.70$ & -2.34 & 6.30 & $19\pm9$ \\
non-dusty SF & 0.020 & $0.7\pm0.3$ & $106\pm 69$ & $ 1.77\pm0.85$ & -3.18 & 5.66 & $25\pm20$ \\
dusty SF & 0.058 & $2.1\pm0.5$ & $188\pm 77$ & $ 3.00\pm0.79$ & -2.41 & 6.19 & $17\pm9$ \\

\hline
\end{tabular}
\begin{tablenotes}
\item[] Uncertainties are from a bootstrap analysis, unless indicated otherwise. Dispersions around the median ($\sigma$) on quantities are normalized median absolute deviations. \citet{chabrier_galactic_2003} IMF assumed for \ms{} and SFRs. SFRs are based on UV luminosity and MIPS 24~$\mu$m fluxes.
\item[a]Statistical and systematic errors, respectively.
\item[b]Excluding galaxies with potential AGN.
\item[c]From fitting to best-fit {\tt EAZY} templates over 1300-1900~\AA{}.
\item[d]Errors reflect Poisson uncertainties. 
\item[e]Quiescent galaxies with MIPS detections were excluded.
\end{tablenotes}
\label{sampletable}
\end{threeparttable}
\end{center}
\end{table*}

The ZFOURGE survey targets 3 legacy fields \citep[CDFS, COSMOS, UDS,][]{giacconi_chandra_2002,scoville_cosmic_2007,lawrence_ukirt_2007} with the near-infrared \fourstar\ camera \citep{persson_fourstar:_2013} on the 6.5m Magellan Baade Telescope.  In each field we obtained a single $\sim11\arcmin\times11\arcmin$ pointing with 5 medium-band filters (\jo, \jt, \jh, \hs, \hl) and the \ks{} filter. Mean limiting $5\sigma$ point-source depths are $25.9$, $25.2$, and $25.2$ mag. in the $J$-bands (FWHM $\sim0.50\arcsec$), $H$-bands ($0.50\arcsec$) and \ks-band ($0.45\arcsec$).  Raw images were processed using our custom pipeline adapted from the one used in the NEWFIRM Medium Band Survey \citep{whitaker_newfirm_2011}.

{\tt SExtractor} \citep{bertin_sextractor:_1996} was used to detect objects in the $Ks-$image and to extract aperture fluxes from PSF-matched versions of the ZFOURGE and other public imaging covering $0.3-8\mu$m. This includes public CANDELS imaging \citep{koekemoer_candels:_2011,grogin_candels:_2011}. The image quality of the Spitzer IRAC+MIPS bands are significantly lower, hence photometry was first deblended using the techniques of \cite{labbe_spitzer_2006} following the prescription detailed in \citet{whitaker_newfirm_2011}.  Our photometric methods are outlined in \citet{tomczak_galaxy_2014} and a full description of the ZFOURGE catalogs will be provided in Straatman et al., (in prep.).

Photometric redshifts and restframe colors are from \eazy{} \citep{brammer_eazy:_2008}. Restframe color uncertainties for individual objects were derived by propagating the full redshift probability distribution in the calculation of the rest-frame color.

Stellar populations were constrained using the \citet{bruzual_stellar_2003} models fitted with {\tt FAST} \citep{kriek_ultra-deep_2009}, assuming exponentially declining star-formation histories (SFR $\sim$ exp[$-t/\tau$], with $\tau$ allowing to vary over log~$[\tau/yr]=7-11$) , Solar metallicity, a \citet{calzetti_dust_2000} dust extinction law (with A$_V=0-4$) and a \citet{chabrier_galactic_2003} initial mass function.

Star-formation rates (SFRs) were computed using restframe $UV$ luminosity computed from the best-fit {\tt EAZY} templates plus the observed $24$~$\mu$m fluxes. Galaxies not detected at $24$~$\mu$m have SFRs that only reflect contribution from the UV. We note the $24$~$\mu$m-based SFRs could suffer systematic uncertainties due to poorly constrained infrared SFR models at restframe $\sim5$~$\mu$m.

In this work, we use three observables to look for signs of active galactic nuclei (AGN): (1) X-ray source catalogs \citep{ueda_subaru/xmm-newton_2008,civano_chandra_2012,xue_chandra_2011}, (2) deep radio source catalogs \citep{schinnerer_vla-cosmos_2010,miller_very_2013}, and (3) excess observed flux at 8~$\mu$m ($3\sigma$ above the best-fit {\tt EAZY} spectral model; e.g. see middle top SED in Fig.~\ref{fig_seds}) plausibly due to hot dust around an AGN. If a galaxy shows any one of these signs it was flagged as a potential AGN. All suspected AGN were excluded from SFR and UV luminosity analysis in this Letter.

\section{The sample}\label{sample}

\begin{figure}
\includegraphics[width=\columnwidth]{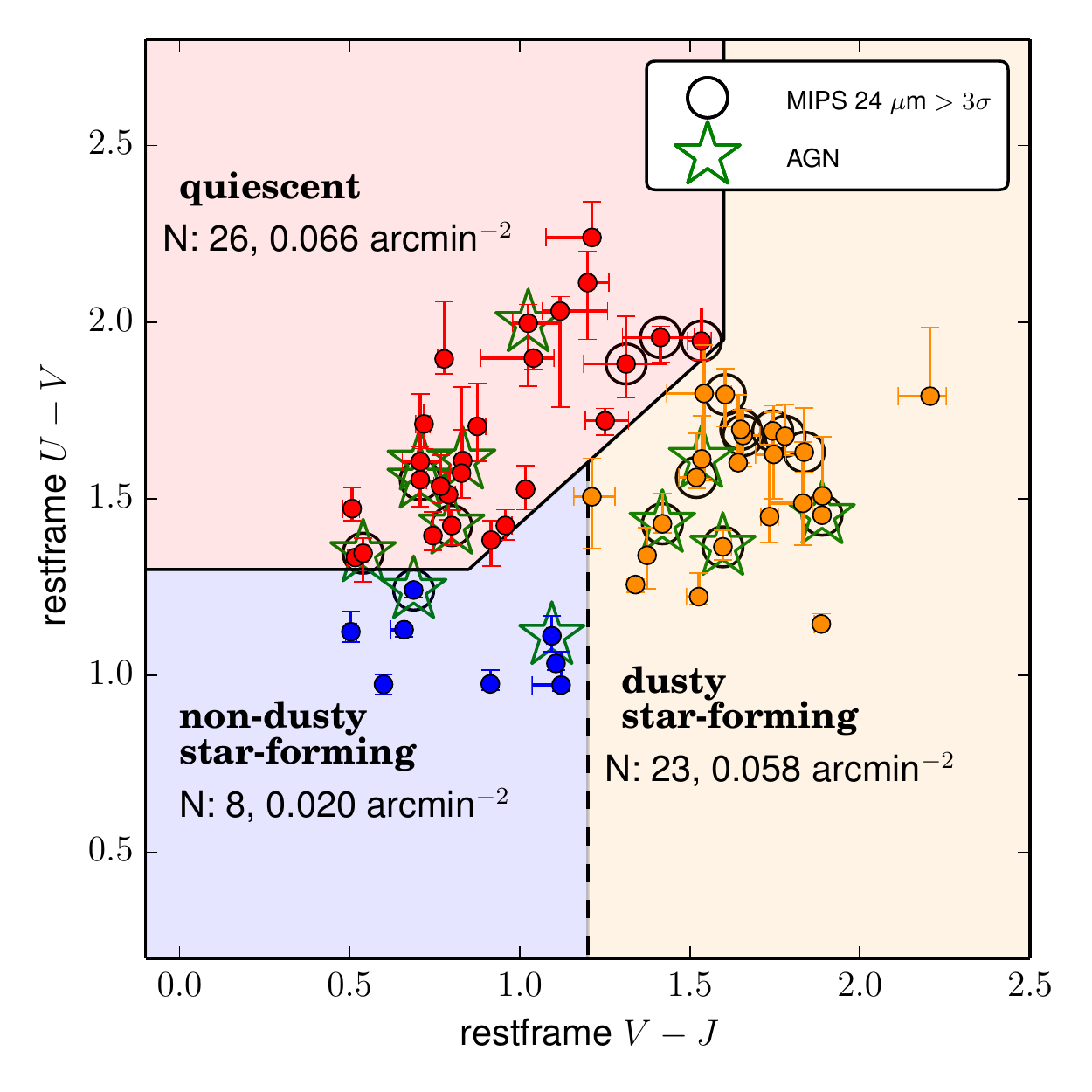}
\caption{
Restframe \uv{} versus \vj{} plot for our mass-limited ZFOURGE sample at $z=3-4$. Following previous work \citep[e.g. ][]{wuyts_what_2007}, we use this information to classify our sample into quiescent and star-forming galaxies, as indicated by the solid-line boundaries. Given the diverse range of restframe colors shown in our sample, we divide the star-forming population into a dusty and non-dusty subgroup, as indicated by the dashed line.
}\label{fig_uvj}
\end{figure}

We constructed a stellar mass limited sample by selecting objects with $>5\sigma$ detections in the \ks-band, medium-band photometric redshifts $z=3-4$, and stellar masses log~\ms{}$>10.6$.  \citet{straatman_substantial_2014} estimate the ZFOURGE survey to be complete to this mass limit at $z\la4$ for single age burst BC03 stellar populations models formed at $z=10$.  We were careful to guard against lower-redshift dusty galaxies with uncertain photometric redshifts scattering into our sample \citep[e.g. ][]{dunlop_systematic_2007}. We therefore exclude 14 galaxies that have a photometric redshift solution $z<3$, when fitted with an extended template including an old stellar population model with large dust attenuation \citep[e.g. ][]{marchesini_most_2010}.  

Our final mass-limited sample at $z=3-4$ contains 57 galaxies (0.14 galaxies arcmin$^{-2}$), about 5\% of the full $z=3-4$ sample (3.0 galaxies arcmin$^{-2}$). The basic properties of our mass-limited sample are provided in Table~\ref{sampletable}.

As shown in the mass-magnitude plots in Fig.~\ref{fig_summary}, at observed visible magnitudes the mass-limited sample (median $R_{tot}=27.48\pm0.41$) is exceedingly faint, fainter even than lower mass ZFOURGE galaxies at $z=3-4$ ($R_{tot}=25.63\pm0.02$).  In \ks{}, the mass-limited sample is relatively bright (median \ks{}$_{tot}=23.32\pm0.14$), implying the sample is much redder (median $R-K=3.87\pm0.19$) than a typical low-mass \ks{}-selected galaxy at $z=3-4$ (median $R-K=1.38\pm0.02$).

Another interesting feature of our mass-limited sample is its intrinsically large spread ($\ga1.5$ mag.) in restframe \uv{} and \vj{} colors (see Fig.~\ref{fig_uvj}). This range is similar to that reported at lower redshifts \citep[e.g. ][]{whitaker_newfirm_2011} and indicates our sample contains a diverse range of SED shapes. 

\section{The diverse properties of $z=3-4$ massive galaxies}\label{galsection}

The large spread in restframe colors prompted us to classify the galaxies based on their restframe colors, following work performed at lower redshift \citep[e.g. ][]{wuyts_what_2007}. As shown in \citet{whitaker_newfirm_2011} galaxies can be divided into two distinct populations at least out to $z\sim2.5$: a quiescent population (red region in Fig.~\ref{fig_uvj}) and a star-forming population.

The fractions of quiescent and star-forming galaxies are \fracqu{} and \fracsa{} with the two error bars reflecting random and systematic errors respectively. The fractions and random uncertainties were calculated by first perturbing the rest-frame colors by their uncertainties and bootstrapping the distribution $10^6$ times. The systematic uncertainties accounts for possible systematics in the rest-frame colors and were obtained by moving \uvj{} color boundary by $\pm0.1$ mag. \citep[e.g. ][]{muzzin_evolution_2013}.

We derived a quiescent fraction independent of the estimated restframe \uvj{} colors by quantifying the fraction of galaxies with low sSFRs. If we define quiescent galaxies to have sSFR~$<0.3$~\igyr{} (see Fig.~\ref{fig_colors}), we derive an alternative quiescent fraction, $0.42\pm0.10$, which agrees with that derived using \uvj{} colors.

We further split the sample into star-forming subgroups containing low and high dust content around $A_V\sim1.6$. We find that redder restframe \vj{} color correlate with larger dust content and a limit of $V$-$J=1.2$ roughly corresponds to a division\footnote{The 3 galaxy classifications are defined using the following vertices:  (\vj, \uv)=($-\infty$,1.3), (0.85,1.3), (1.2, 1.6), (1.6,1.95), (1.6,$+\infty$)} around $A_V\sim1.6$ (for constant star-formation rate models and age $<1$ Gyr). Henceforth, we refer to the blue (restframe $V$-$J<1.2$) star-forming galaxies as relatively unobscured and red (restframe $V$-$J>1.2$) star-forming galaxies as dusty. Traditional Lyman break samples generally do not show higher $A_V$ values, presumably due to reduced selection efficiency of redder galaxies \citep[e.g. ][]{papovich_stellar_2001}. The unobscured and dusty star-forming subgroup fractions are: \fracsf{}, and \fracdu{}, respectively.

\begin{figure*}
\includegraphics[width=\textwidth]{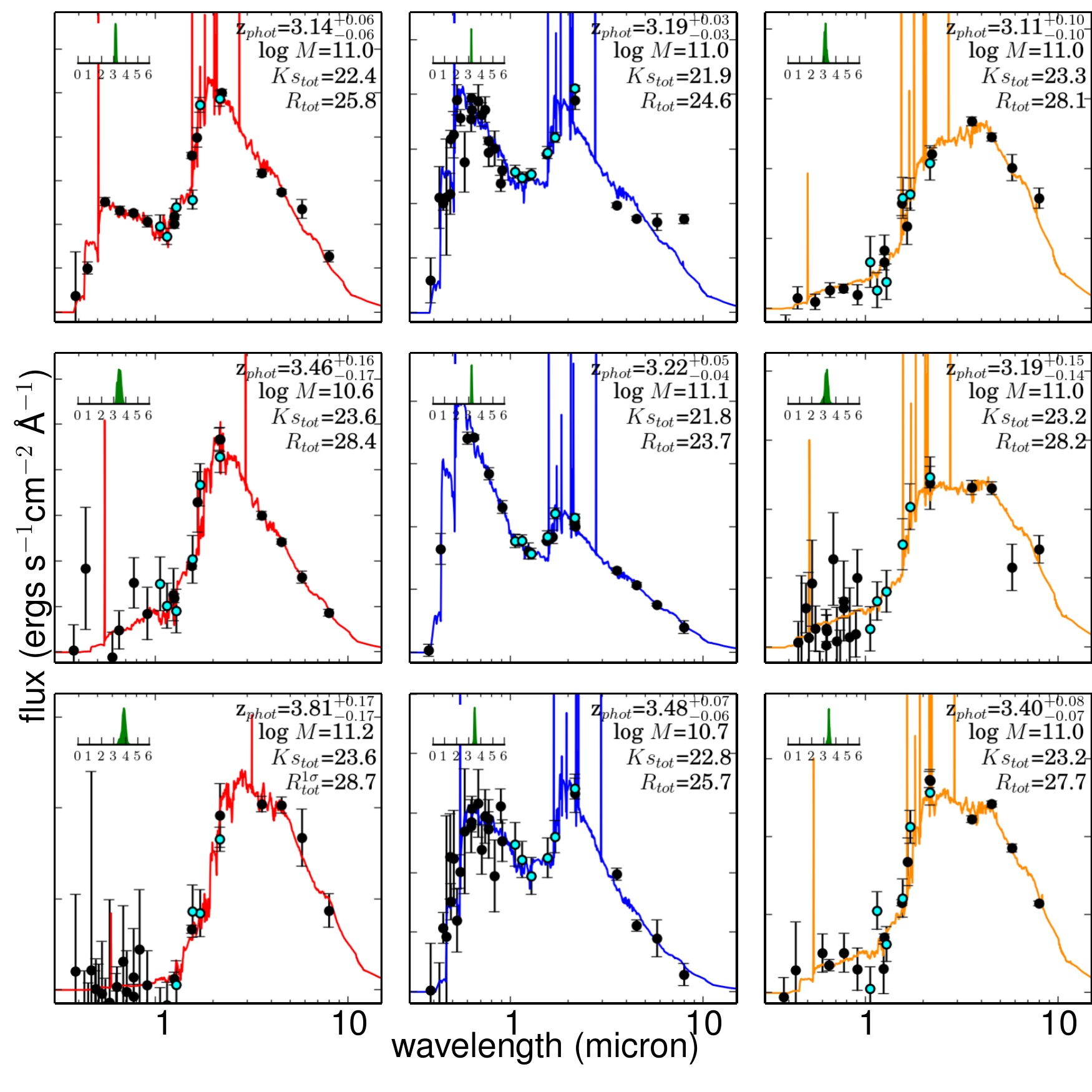}
\caption{Example observed galaxy SEDs for each galaxy subgroup identified in Fig.~\ref{fig_uvj}: quiescent, star-forming, and dusty star-forming from left to right. Spectra are the best-fit {\tt EAZY} photometric redshift models, black points are from public imaging and cyan points are from ZFOURGE imaging. Insets show the photometric redshift probability distributions from {\tt EAZY}.
\label{fig_seds}}
\end{figure*}

Some example SED fits for each galaxy subgroup are given in Fig.~\ref{fig_seds}. We review subgroups properties here:

\begin{itemize}
\item {\bf Quiescent galaxies} (red in all figures) make up \numqu{}/\numgals{} of our sample and their SEDs (see Fig.~\ref{fig_seds}) have sharp Balmer breaks, indicating the stellar populations are dominated by old stars and suppressed star-formation rates \citep[e.g. ][]{straatman_substantial_2014}.  Fig.~\ref{fig_uvj} shows that the majority (20/26) of our quiescent population are not detected at $24$~$\mu$m and therefore have low SFR values, with a median sSFR of $0.14\pm0.03$~\igyr{} for $24$~$\mu$m non-detections. Their median age is $1.0\pm0.1\times10^9$~yr, which is longer than their median $\tau$ of $1.6\pm0.4\times10^8$~yr, indicating that these galaxies do not contain sizable young stellar populations.

A small fraction (6/\numqu{}) of the quiescent galaxies have MIPS detections and high implied SFRs. This is surprising given the old ages implied by their SEDs. Three of the MIPS detected sources show signs of AGN, which may explain the MIPS fluxes. The remaining 3 fall near the \uvj{} boundary between dusty star-forming and quiescent galaxies and might have scattered into the quiescent region.

\item {\bf The entire star-forming population} shows a median SFR of $180\pm60$ and a sSFR of $2.7\pm0.7$. The median $A_V$ is $1.7\pm0.3$, so the average massive star-forming galaxy is dusty (see Fig.~\ref{fig_uvj}).
	
\begin{itemize}
	\item {\bf Star-forming galaxies with little dust} (A$_V=1.0\pm0.1$; blue in all figures) only make up \numsf{}/\numgals{} of our sample and show lower stellar masses compared to the whole sample (median log \ms{} $=10.66\pm0.08$ \msun{}). Their median SFR is $110\pm70$~\msunyear{} and they have a median sSFR of $2.0\pm0.9$~\igyr{}. With median UV luminosities of $M_{1700}=-20.53\pm0.28$, they are $\sim5$ mag. fainter than UV-luminous $B-$dropout galaxies in the UV \citep{lee_average_2011} as shown in Fig.~\ref{fig_colors}.

	\item The {\bf dusty, star-forming galaxies } (orange in all figures) constitute \numdu{}/\numgals{} of our sample and show a roughly similar \ms{} distribution to the quiescent population (see Table~\ref{sampletable} and Fig.~\ref{fig_summary}). As shown in the SFR plot of Fig.~\ref{fig_colors}, the dusty subgroup shows a median SFR of $190\pm80$~\msunyear{}. The median sSFR of this subgroup is sSFR~$=3.0\pm0.8$~\igyr{}. Restframe optical fluxes appear to be significantly obscured due to dust, with best-fit {\tt FAST} A$_V$ values ranging from $\sim1-3$ and a subgroup median of A$_V=2.0\pm0.2$. 
\end{itemize}

\end{itemize}

\section{Comparison to high-redshift samples}\label{highzsection}

\subsection{\bbdrop{} samples}\label{dropoutsection}

As most galaxy studies at $z>3$ have relied on Lyman break dropout selection techniques, we briefly compare our mass-limited sample to \bbdrop{} samples \citep[e.g. ][]{giavalisco_rest-frame_2004}, which selects galaxies over a similar redshift range \citep[e.g.~$3.2<z<4.4$, ][]{oesch_rest-frame_2013}.

By integrating \citet{lee_how_2012} UV-selected stellar mass functions to our mass limit, we estimate $20\pm14$ \bdrop{} galaxies should be found in our fields. This is lower than what we find, $\numgals{}\pm11$ (errors Poisson plus field-to-field variance), but is more consistent with our unobscured star-forming subgroup (\numsf{}).

\begin{figure*}
\includegraphics[width=\textwidth]{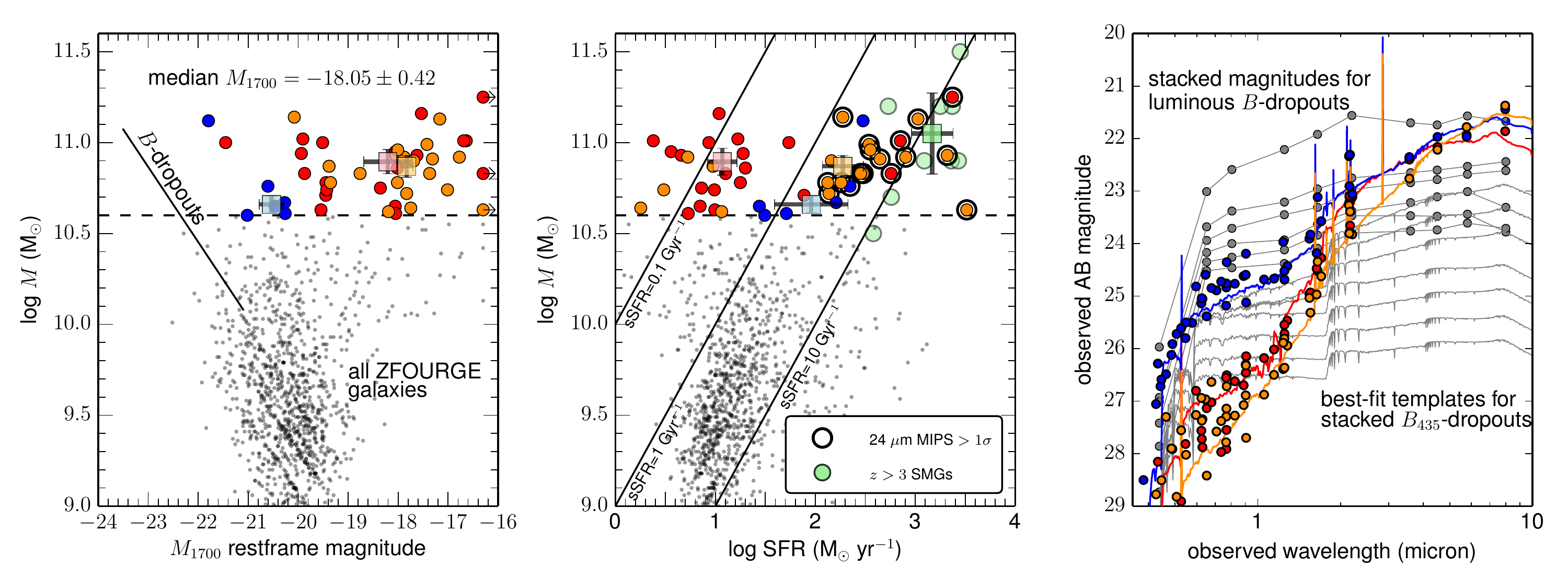}
\caption{
{\it Left panel:} Stellar mass versus UV luminosity for our mass-limited sample (colored points).  Grey points are $z=3-4$ low-mass ZFOURGE galaxies. Median values are provided for each subgroup as shaded squares.  The restframe UV luminosity of our mass-limited sample is fainter than expected from a reported correlation of very luminous $B$-dropout galaxies \citep{lee_average_2011}. 
{\it Center panel:} Mass-SFR plot with light green circles representing submillimeter galaxies compiled by \citet{toft_submillimeter_2014}. These have $4\times$ higher median specific SFRs compared to dusty galaxies in our sample.
{\it Right panel:} Observed stacked SEDs for the mass-limited $z=3-4$ ZFOURGE galaxies. Each point reflects the median flux for each bandpass. Spectra are median stacked best-fit {\tt EAZY} spectral templates after shifting to the median redshift of each subgroup (see Table~\ref{sampletable}). These SEDs are compared to the \citet{oesch_rest-frame_2013} \bdrop{} sample, whose stacked galaxy SEDs (binned by $M_z$ restframe) are best-fit by the templates shown in grey.  Grey circles are stacked magnitudes from the more luminous \bbdrop{} galaxy sample (binned by $I$-band observed) of \citet{lee_average_2011}. 
\label{fig_colors}}
\end{figure*}

It has been suggested in previous works that UV luminosity correlates with stellar mass \citep[e.g. ][]{lee_average_2011}. The existence of such a correlation would be important, as it implies we could use UV luminosity as a proxy for stellar mass. However, as shown in Fig.~\ref{fig_colors}, our sample shows no such trend.  On the contrary, galaxies with log~\ms{}$>10.6$ have median \muv{}, are 2 mag {\em fainter} than log~\ms{}$=10.0$ galaxies, and are 5 mag fainter than expected from the reported \ms{}$-M_{1700}$ correlation for \bbdrop{} \citep{lee_average_2011}.

In Fig.~\ref{fig_colors} the SEDs of our mass-limited galaxies are compared to dropout galaxies \citep{lee_average_2011,oesch_rest-frame_2013}. At the restframe UV, the ZFOURGE dusty and quiescent SEDs are fainter than the faintest dropout galaxies shown and rapidly rise in luminosity to brighter magnitudes at longer wavelengths.

Overall we find that the SEDs of our mass-limited sample are much redder at both restframe UV and optical compared to dropout galaxies at similar redshifts.

\subsection{Sub-mm detected samples}\label{firsection}

How do our dusty galaxies relate to dusty starbursts selected at submillimeter wavelengths? None of our galaxies have a secure detection in public submillimeter catalogs \citep{aretxaga_aztec_2011,hodge_alma_2013}.  This is perhaps not surprising given the $8\times$ lower SFRs ($190\pm80$ \msunyear{}) and $5\times$ higher surface number density ($210\pm40$~deg.$^{-2}$) compared to bright ($F_{1.1mm}\ga4.2$ mJy) SMGs at redshifts $z=3-6$ \citep{toft_submillimeter_2014}.  

Fig.~\ref{fig_colors} provides a direct comparison between our sample and a $z=3-6$ SMG compilation from \citet{toft_submillimeter_2014}. The median sSFR for the dusty galaxies (sSFR$=3.0\pm0.8$~\igyr{}) is much lower than the SMGs ($11.7\pm0.8$~\igyr{}) at the same redshift, suggesting our dusty galaxies are in a less extreme, more typical mode of star-formation.

\section{Summary and Conclusions}\label{discussion}

Using new data from the ZFOURGE survey we have isolated a high-quality sample of \numgals{} galaxies at $z=3-4$. Our sample is complete to a stellar mass of log~\ms{}~$>10.6$ and provides the first census of $M^{\star}$ galaxies at these redshifts.  Our results benefit significantly from deep photometry in medium-band near-IR filters that yield robust photometric redshifts and robust restframe colors. 

Remarkably, the average massive galaxy is extremely faint in the observed optical $R=27.48\pm0.41$ AB, and very red. They are fainter in the rest-frame UV compared to less massive galaxies and we find no evidence for a relation between UV-luminosity and stellar mass.

We find strong evidence that the massive galaxy population at $z=3-4$ spans a diverse range in stellar age and dust content, as suggested by earlier shallower surveys \citep[e.g. ][]{marchesini_most_2010} and similar to massive galaxies lower redshift \citep[e.g. ][]{van_dokkum_space_2006,muzzin_evolution_2013}. The quiescent fraction is very high (\percqu{}) suggesting that quenching was already very efficient at early times \citep[see also ][]{straatman_substantial_2014}.

The star forming galaxies (\percsa{} of the massive sample) are on average rather dusty ($A_V\sim1.7\pm0.3$) but span a large range in rest-frame colors. Subdividing star forming galaxies by their rest-frame \vj{} colors in red ``dusty'' (\percdu{}) and blue ``non-dusty'' galaxies (\percsf{}) shows that $\sim86\%$ of massive galaxies at these redshifts are characterized by very red colors. These colors are much redder than those of similarly massive Lyman break galaxies \citep{lee_average_2011}.

Dusty star-forming galaxies make up a significant portion of our sample and dominate the star-forming galaxies ($74\%$). Prior to our work, this type of galaxy has not been securely selected at restframe optical because with typical $R\sim28$ and \ks{}~$\sim24$ magnitudes (see Fig.~\ref{fig_summary}) they are fainter than, or are just at the limits of, previous \ks{} selected samples. Compared to typical \bbdrop{} galaxies, the massive dusty $z=3-4$ galaxies show higher SFRs \citep[e.g.][]{smit_star_2012}, but contribute only $\sim10\%$ to the total star formation rate density with log~($\rho_{SFR}$/(\msunyear{}~Mpc$^{-3}$))~$=-2.41$ \citep{bouwens_uv-continuum_2012}.

It is well known that there exists a population of dust obscured extreme star forming galaxies at high redshift, selected in the sub-millimeter \citep{smail_nature_2002}. These are thought to be the ``tip of the iceberg'' of dust-obscured star forming galaxies. Here we uncover a reservoir of highly obscured star-forming galaxies with similar stellar masses, but lower SFRs. Overall, massive dusty galaxies have $4\times$ lower specific star-formation rates and are $\sim5\times$ more numerous compared to $z>3$ submillimeter galaxies.

At face value, this would seem to suggest the dusty galaxies are undergoing a more typical mode of massive galaxy star-formation compared to the rare, extreme star-bursts associated with submillimeter galaxies.  Future, deeper (sub)millimeter surveys with SCUBA2 and ALMA will provide direct evidence for the link between these populations. Clearly, spectroscopic follow up and confirmation of the redshifts of these dusty galaxies is high priority, but will be very challenging.\\

To conclude, we now have access to all galaxy stellar population types at $z=3-4$: from the least obscured Lyman break galaxies to heavily obscured galaxies, as well as quiescent galaxies.  By linking these galaxy populations to massive galaxies at lower redshifts we will be able to develop a more complete picture of massive galaxy formation over the past 12 billion years.

\acknowledgments We thank the reviewer for providing useful comments that greatly improved the text. We thank Pascal Oesch for providing data from his work.  Australian access to the Magellan Telescopes was supported through the National Collaborative Research Infrastructure Strategy of the Australian Federal Government. LRS and KG acknowledge funding from a Australian Research Council Discovery Program grant DP1094370. IL acknowledges support from ERC HIGHZ \#227749 and NL-NWO Spinoza. CP, KVT and VT acknowledge support from National Science Foundation grant AST-1009707.

{\it Facilities:}\facility{Magellan(\fourstar)}.

\bibliographystyle{apj}

\end{document}